\documentclass[aps,pra,twocolumn,reprint]{revtex4-2}

\usepackage{graphicx}
\usepackage{amsmath, amssymb, amsfonts}
\usepackage{bm}
\usepackage{xcolor}
\usepackage[normalem]{ulem}

\DeclareMathAlphabet\mathbfcal{OMS}{cmsy}{b}{n}

\newlength{\figwidth}
\newlength{\lfig}
\newlength{\sfig}
\begin{document}
\title{Making molecules by mergoassociation: the role of center-of-mass motion}

\author{Robert C. Bird}
\affiliation{Joint Quantum Centre (JQC) Durham-Newcastle, Department of
Chemistry, Durham University, South Road, Durham, DH1 3LE, United Kingdom.}
\author{Jeremy M. Hutson}
\email{j.m.hutson@durham.ac.uk} \affiliation{Joint Quantum Centre (JQC)
Durham-Newcastle, Department of Chemistry, Durham University, South Road,
Durham, DH1 3LE, United Kingdom.}

\date{\today}

\date{\today}

\begin{abstract}
	In \emph{mergoassociation}, two atoms in separate optical traps are combined to form a molecule when the traps are merged. Previous theoretical treatments have considered only the relative motion of the atoms, neglecting coupling to the motion of the center of mass. We develop a theoretical method to include the coupling to center-of-mass motion and consider its consequences for experiments for both weak and strong coupling. We consider the example of RbCs and then extend the treatment to other systems where mergoassociation may be effective, namely RbSr, RbYb and CsYb. We consider the role of the coupling when the traps are anisotropic and the potential use of moveable traps to construct quantum logic gates.
\end{abstract}

\maketitle

\section{Introduction}

\begin{figure}[t]
\begin{center}
\includegraphics[width=0.43\textwidth]{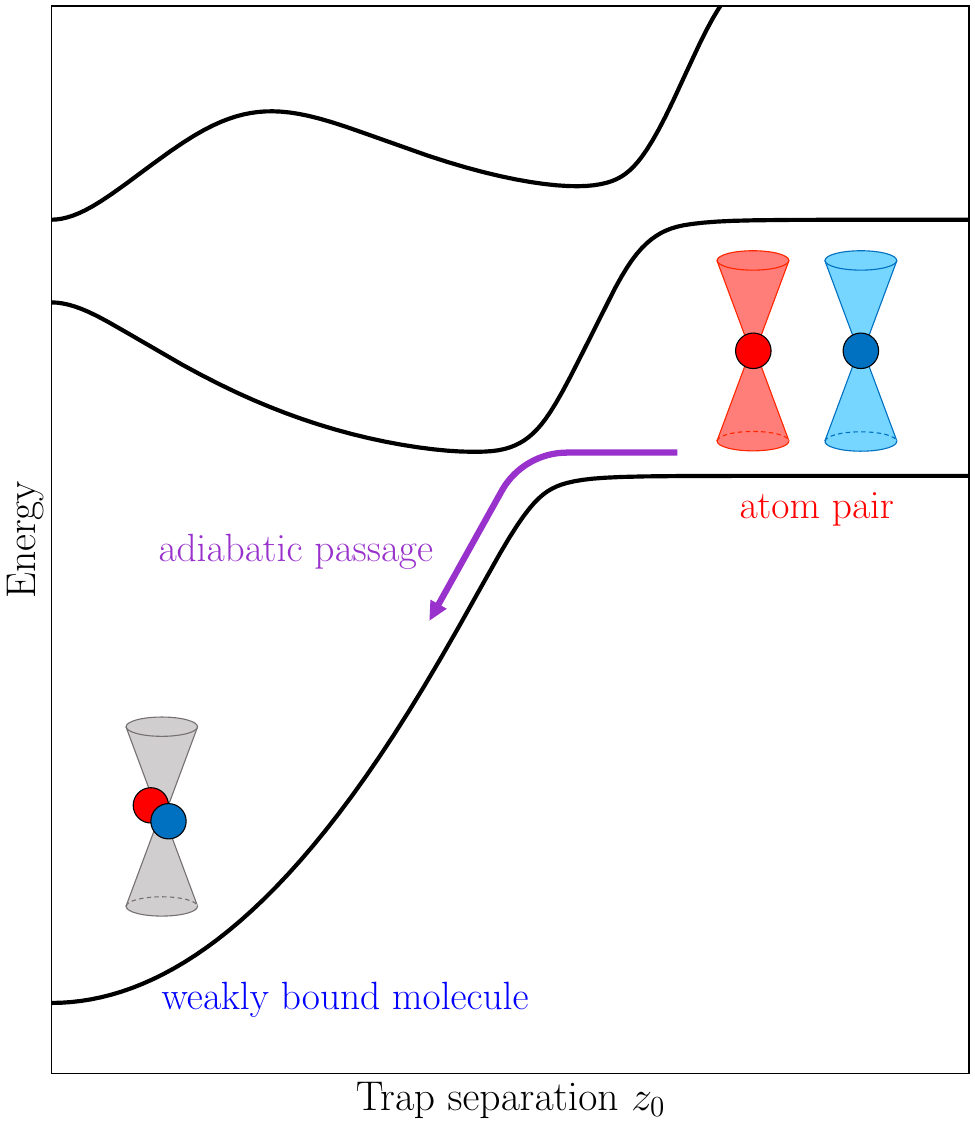}
	\caption{Schematic representation of the energy levels involved in mergoassociation, as a function of trap separation $z_0$. The molecular level (approximately quadratic as a function of $z_0$) has avoided crossings with motional states of the atom pair (approximately horizontal at large $z_0$). Mergoassociation occurs when an atom pair in the lowest motional state is transferred into the molecular state by adiabatic passage over the lowest avoided crossing. The levels shown neglect motion of the center of mass.
\label{fig:intro}
}
\end{center}
\end{figure}

Recent experiments \cite{Ruttley:2023} have shown that two ultracold atoms, confined in separate optical traps or tweezers, may combine to form a weakly bound molecule when the traps are merged. The process occurs because the energies of high-lying molecular states cross the energy of the atom pair as a function of trap separation. Coupling between the atom-pair and molecular states generates an avoided crossing between the states. Atom pairs can thus be converted into molecules by adiabatic passage as the traps are merged. The process is known as \emph{mergoassociation} and has great potential for creating ultracold molecules that are inaccessible with other methods. The levels involved are shown schematically in Fig.\ \ref{fig:intro}.

The levels produced when two traps merge were first studied by Stock \emph{et al.}\ \cite{Stock:2003, Stock:2006}. They considered two atoms that are identically trapped. Under these circumstances, with harmonic traps, there is an exact separation of the motions in the relative and center-of-mass coordinates. Their calculations dealt entirely with the relative motion and with spherical traps. Following the experimental work \cite{Ruttley:2023}, which was carried out with Rb and Cs atoms in nonidentical optical tweezers with large anisotropy, we extended the formal theory to handle nonidentical, anisotropic traps \cite{Bird:mergo:2023}. However, the numerical calculations presented in ref.\ \cite{Bird:mergo:2023} were still limited to the relative motion, neglecting coupling to the motion of the center of mass.

The purpose of the present paper is to investigate the influence of center-of-mass motion on the energy levels involved in mergoassociation and to consider their implications for experiments. Idziaszek \emph{et al.}\ \cite{Idziaszek:2007} briefly considered the coupling between the relative and center-of-mass motions for atom-ion interactions in one dimension and commented that certain avoided crossings were weaker. However, the problem has not been considered in 3 dimensions and the dependence of the level patterns on the coupling strength has not been explored. The influence of coupling between the relative and center-of-mass motions on the levels that arise for two interacting particles in a single trap has been studied more extensively \cite{Deuretzbacher:2008, Mentink:2009, Jachymski:trap:2013, Jachymski:2016, Jachymski:2020, Hood:NaCs:2020}.

The structure of this paper is as follows. Section \ref{sec:theory} introduces the problem and describes the methods we use, including an important modification of the molecular basis functions that dramatically improves convergence. Section \ref{sec:effects} uses the example of RbCs to explore the effects of the coupling between relative and center-of-mass motions for both weak and strong coupling (Sections \ref{sec:weak} and \ref{sec:strong}). We consider the consequences of the coupling for mergoassociation starting from atoms either in their motional ground states or in motionally excited states. This section also explores mergoassociation for other systems, considering the examples of RbSr, RbYb and CsYb (Section \ref{sec:other-systems}), the effect of the strong anisotropy of the tweezer traps used in current experiments (Section \ref{sec:aniso}), and the potential use of moveable traps to construct quantum logic gates (Section \ref{sec:logic}). Finally, Section \ref{sec:conclusions} presents our conclusions.

\section{Theoretical Methods}
\label{sec:theory}

We consider two atoms independently confined in adjacent optical traps. Atom $i$ has mass $m_i$ and position $\boldsymbol{R}_i$ and is confined in a trap centered at $\boldsymbol{R}_i^0$. The motion may be factorized approximately into terms involving the relative and center-of-mass coordinates of the pair, $\boldsymbol{R}$ and $\mathbfcal{R}$ respectively. The 2-atom kinetic-energy operator is exactly separable,
\begin{align}
-\frac{\hbar^2}{2m_1}\nabla_1^2 - \frac{\hbar^2}{2m_2}\nabla_2^2
&= -\frac{\hbar^2}{2\mu}\nabla_R^2 -\frac{\hbar^2}{2{\cal M}}\nabla_{\cal R}^2 \nonumber\\
&= \hat{T}_\textrm{rel} + \hat{T}_\textrm{com},
\end{align}
where
\begin{align}
\mathbfcal{R}   &= \left(m_1 \boldsymbol{R}_1   + m_2 \boldsymbol{R}_2  \right)/{\cal M};\\
\boldsymbol{R}   &= \boldsymbol{R}_2   - \boldsymbol{R}_1;\\
{\cal M} &= m_1+m_2;\\
\mu &= m_1 m_2 / {\cal M}.
\end{align}

If the individual traps are harmonic but non-spherical, their combined potential energy is
\begin{align}
V^\textrm{trap} &= \sum_i \textstyle{\frac{1}{2}} m_i [\boldsymbol{R}_i-\boldsymbol{R}_i^0]^\intercal \boldsymbol{\omega}_i^2 [\boldsymbol{R}_i-\boldsymbol{R}_i^0],
\label{eq:unsep}
\end{align}
where $\boldsymbol{\omega}_1$ and $\boldsymbol{\omega}_2$ are second-rank tensors of trap frequencies. Equation \ref{eq:unsep} may be rearranged to \cite{Bird:mergo:2023}
\begin{align}
V^\textrm{trap} =\ & V^\textrm{trap}_\textrm{rel}(\boldsymbol{R}) + V^\textrm{trap}_\textrm{com}(\mathbfcal{R}) + V^\textrm{trap}_\textrm{cpl}(\boldsymbol{R},\mathbfcal{R}) \nonumber\\
=\ & \textstyle{\frac{1}{2}} \mu [\boldsymbol{R}-\boldsymbol{R}_0]^\intercal \boldsymbol{\omega}_\textrm{rel}^2 [\boldsymbol{R}-\boldsymbol{R}_0] \nonumber\\
&+\textstyle{\frac{1}{2}} {\cal M} [\mathbfcal{R}-\mathbfcal{R}_0]^\intercal \boldsymbol{\omega}_\textrm{com}^2 [\mathbfcal{R}-\mathbfcal{R}_0] \nonumber\\
&+ \mu [\mathbfcal{R}-\mathbfcal{R}_0]^\intercal \Delta\boldsymbol{\omega}^2 [\boldsymbol{R}-\boldsymbol{R}_0],
\label{eq:sep-nonspher}
\end{align}
where
\begin{align}
\boldsymbol{R}_0 &= \boldsymbol{R}_2^0 - \boldsymbol{R}_1^0;\\
\boldsymbol{\omega}_\textrm{rel}^2 &= \left({m_2\boldsymbol{\omega}_1^2+m_1\boldsymbol{\omega}_2^2}\right)\big/{\cal M};
\label{eq:wrel}\\
\mathbfcal{R}_0 &= \left(m_1 \boldsymbol{R}_1^0 + m_2 \boldsymbol{R}_2^0\right)/{\cal M};\\
\boldsymbol{\omega}_\textrm{com}^2 &= \left(m_1\boldsymbol{\omega}_1^2+m_2\boldsymbol{\omega}_2^2\right)/{\cal M};\\
\Delta\boldsymbol{\omega}^2 &= \boldsymbol{\omega}_2^2 - \boldsymbol{\omega}_1^2.
\label{eq:conf-freq}
\end{align}
The last term in Eq.\ \ref{eq:sep-nonspher} is the motional coupling between relative and center-of-mass motions, characterized by $\Delta\boldsymbol{\omega}^2$.

We restrict the discussion here to the case where the two traps are coaligned, so that the tensors $\boldsymbol{\omega}_1^2$, $\boldsymbol{\omega}_2^2$, $\boldsymbol{\omega}_\textrm{rel}^2$, $\boldsymbol{\omega}_\textrm{com}^2$ and $\Delta\boldsymbol{\omega}^2$ all have the same principal axes
and $\Delta\boldsymbol{\omega}^2 = (\boldsymbol{\omega}_1+\boldsymbol{\omega}_2) (\boldsymbol{\omega}_2-\boldsymbol{\omega}_1).$
We choose Cartesian axes along these principal axes, so that the tensors are all diagonal.
$\boldsymbol{R}$, $\boldsymbol{R}_0$, $\mathbfcal{R}$ and $\mathbfcal{R}_0$ are column vectors; the components of $\boldsymbol{R}$ and $\mathbfcal{R}$ are denoted $x$, $y$, $z$ and $X$, $Y$, $Z$, respectively, and similarly for $\boldsymbol{R}_0$ and $\mathbfcal{R}_0$.

In ref.\ \cite{Bird:mergo:2023} we developed a basis-set approach that gives accurate results for the energy levels of relative motion for separated traps. This uses a nonorthogonal basis set made up of 3-dimensional harmonic-oscillator functions centered at $\boldsymbol{R}=\boldsymbol{R}_0$, supplemented with a single function $\psi_a$ for the molecular state.
The Hamiltonian for relative motion may be written
\begin{align}
\hat{H}_\textrm{rel}
&= \hat{T}_\textrm{rel} + V_\textrm{rel}^\textrm{trap}(\boldsymbol{R}) + V_\textrm{int}(\boldsymbol{R}) \nonumber\\
&= \hat{H}_\textrm{rel}^\textrm{trap} + V_\textrm{int}(\boldsymbol{R})
= \hat{H}_\textrm{int} + V_\textrm{rel}^\textrm{trap}(\boldsymbol{R}),
\end{align}
where $\hat{H}_\textrm{rel}^\textrm{trap}$ is the Hamiltonian for the nonspherical harmonic trap and $\hat{H}_\textrm{int}$ is the Hamiltonian for the untrapped atom pair. If $V_\textrm{int}(\boldsymbol{R})$ is represented as a contact potential at the origin \cite{Huang:1957} that corresponds to scattering length $a>0$, $\hat{H}_\textrm{int}$ has a single molecular bound state, with eigenfunction
\begin{equation}
\psi_a = (2\pi a)^{-1/2} R^{-1} \exp(-R/a),
\label{eq:int-fun}
\end{equation}
and eigenvalue
\begin{equation}
E_a = -\hbar^2/(2\mu a^2).
\label{eq:int-e}
\end{equation}
The elements of the Hamiltonian and overlap matrices for relative motion are summarized in the Appendix.

Here we extend this approach to take account of motion in the center-of-mass coordinate $\mathbfcal{R}$.
The full Hamiltonian is
\begin{equation}
\hat{H} = \hat{H}_\textrm{rel} + \hat{H}^\textrm{trap}_\textrm{com}(\mathbfcal{R}) + V^\textrm{trap}_\textrm{cpl}(\boldsymbol{R},\mathbfcal{R}),
\end{equation}
where $\hat{H}_\textrm{com}^\textrm{trap}(\mathbfcal{R})=\hat{T}_\textrm{com}+V^\textrm{trap}_\textrm{com}(\mathbfcal{R})$.

\subsection{Direct-product approach}
\label{sec:direct}

The simplest approach is to multiply each function in the basis set for relative motion with a set of 3-dimensional harmonic-oscillator functions in the center-of-mass coordinate.
The harmonic functions are all eigenfunctions of $\hat{H}_\textrm{com}^\textrm{trap}(\mathbfcal{R})$, which are centered at $\mathbfcal{R}=\mathbfcal{R}_0$. The resulting direct-product functions are represented by Dirac kets $|n_x n_y n_z N_X N_Y N_Z\rangle$ or $|a N_X N_Y N_Z\rangle$, and the resulting matrix elements are given in the Appendix. For spherical traps or traps displaced along $z$, the basis set may be factorized into 4 symmetry blocks with $n_x+N_x$ and $n_y+N_Y$ either even (E) or odd (O), and calculations are carried out for each block separately.

\begin{figure}[tbp]
	\includegraphics[width=0.43\textwidth]{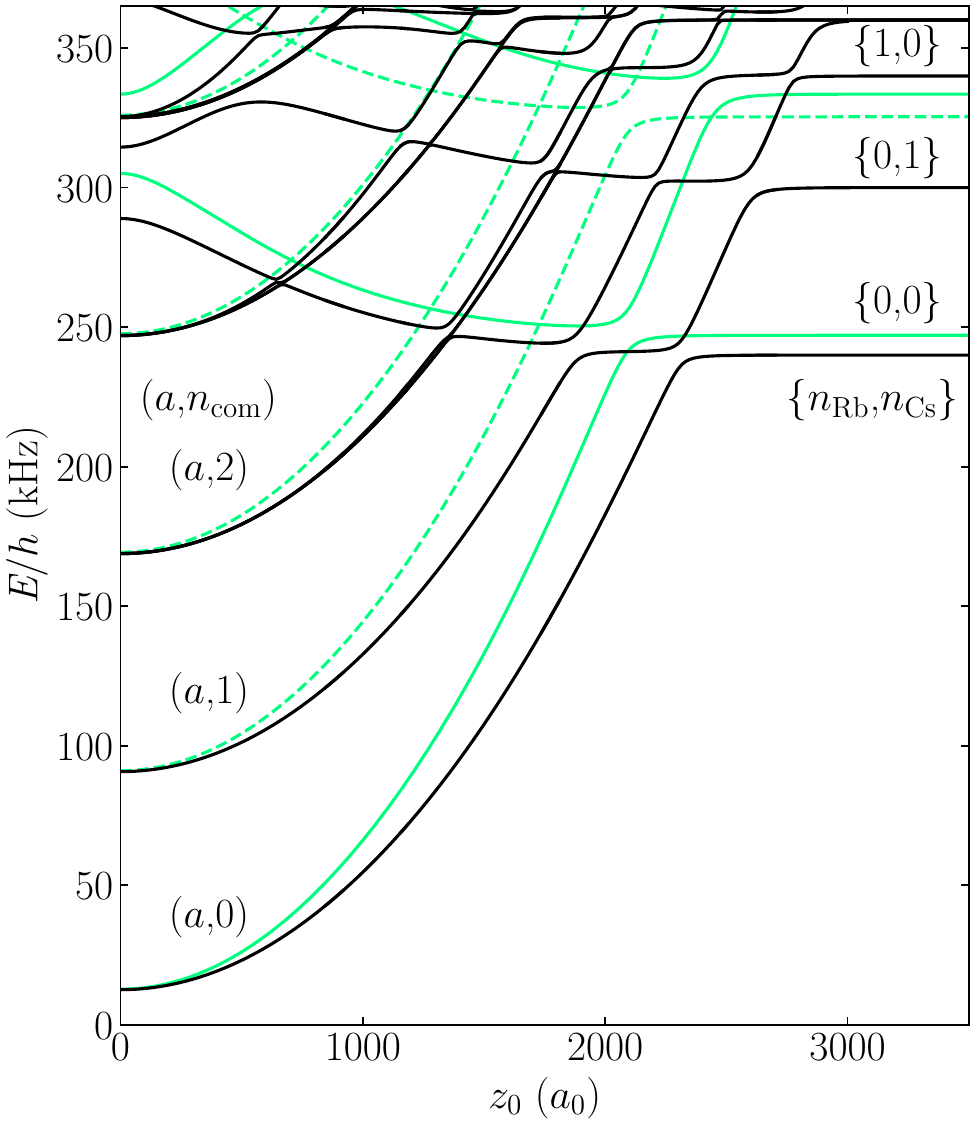}
	\caption{Levels of Rb and Cs atoms in separated spherical traps as a function of separation $z_0$, with $\omega_\textrm{Rb}=100$~kHz and $\omega_\textrm{Cs}=60$~kHz. Solid green lines show the levels for pure relative motion, while dashed green lines show levels excited in the center-of-mass coordinate but neglecting coupling between relative and center-of-mass motions. Black lines show the results of the full coupled calculation using a direct-product basis set (444)(444). Only levels with EE symmetry are shown.}
\label{fig:direct-prod}
\end{figure}

Figure \ref{fig:direct-prod} shows an example of energy levels for Rb and Cs in separated spherical traps as a function of $z_0$. These are calculated with a scattering length $a=554\ a_0$; this corresponds to a bound-state energy $E_a/h \approx -112$ kHz, suitable for RbCs at the magnetic field used for mergoassociation in ref.\ \cite{Ruttley:2023}. The black lines are obtained with a large direct-product basis set with $(n_x^\textrm{max} n_y^\textrm{max} n_z^\textrm{max}) (N_X^\textrm{max} N_Y^\textrm{max} N_Z^\textrm{max}) = (444)(444)$. This basis set contains 4270 functions for EE symmetry. The near-horizontal levels are those of pairs of trapped atoms that at large separation are in separate traps; they show single-atom trap excitations of frequency 60 and 100~kHz. They are labeled by the principal quantum numbers $\{n_\textrm{Rb},n_\textrm{Cs}\}$ of the individual 3d harmonic traps. Their wavefunctions are not simply expressed in terms of relative and center-of-mass motions. The levels that vary quadratically with $z_0$ are molecular states and are labeled $(a,n_\textrm{com})$, where $n_\textrm{com}$ is the principal quantum number for center-of-mass motion. These two sets of levels undergo avoided crossings with one another.

Figure \ref{fig:direct-prod} compares these results with an approximation (green lines) that neglects the coupling $V^\textrm{trap}_\textrm{cpl}(\boldsymbol{R},\mathbfcal{R})$ between relative and center-of-mass motions. The uncoupled levels for the ground state of center-of-mass motion are shown as solid green lines, with levels excited in center-of-mass motion parallel to them and shown as dashed green lines. In this approximation, the atom-pair levels have incorrect energies governed by $\omega_\textrm{rel}$ and $\omega_\textrm{com}$. In addition, the uncoupled molecular levels are shifted upwards from the uncoupled ones by an amount that varies with $z_0$.

\subsection{Shifted-molecule approach}
\label{sec:shift-mol}

The direct-product basis set has the disadvantage that there are non-zero matrix elements of the form
\begin{equation}
\langle a N'_X N'_Y N'_Z|V^\textrm{trap}_\textrm{cpl}(\boldsymbol{R},\mathbfcal{R})|a N_X N_Y N_Z\rangle.
\end{equation}
These matrix elements are diagonal in $a$ but off-diagonal in $N_X$, $N_Y$ or $N_Z$ by 1 when the trap separation $\boldsymbol{R}_0$ has components along $X$, $Y$ or $Z$, respectively. They are due to the term $\mu [\mathbfcal{R}-\mathbfcal{R}_0]^\intercal \Delta\boldsymbol{\omega}^2 \boldsymbol{R}_0$ in Eq.\ \ref{eq:sep-nonspher}, which shifts the minimum in the potential for center-of-mass motion away from $\mathbfcal{R}_0$ for molecular states. As a result, convergence with respect to the basis set for center-of-mass motion is poor when $\Delta\boldsymbol{\omega}^2 \boldsymbol{R}_0$ is substantial.

To circumvent this issue, we use a modified basis set where the functions for motion in $\mathbfcal{R}$ are shifted for the molecular state. They are still harmonic-oscillator functions with the same frequency, but are centered at $\tilde{\mathbfcal{R}}_0 = \mathbfcal{R}_0 - \Delta\mathbfcal{R}$, where
\begin{equation}
\Delta\mathbfcal{R} = \frac{\mu}{M} [\boldsymbol{\omega}_\textrm{com}^2]^{-1} \Delta\boldsymbol{\omega}^2 \boldsymbol{R}_0.
\end{equation}
The resulting shifted-molecule functions are represented by Dirac kets $|a \tilde{N}_X \tilde{N}_Y \tilde{N}_Z\rangle$. The kets $|n_x n_y n_z N_X N_Y N_Z\rangle$ are retained unmodified, centered on $\mathbfcal{R}_0$. The matrix elements in the shifted-molecule basis set are given in the Appendix.

The most important effect of the shifted-molecule basis set is that the diagonal matrix elements for all molecular functions are shifted in energy by
\begin{equation}
\Delta E_a = -\frac{\mu}{2} \boldsymbol{R}_0^\intercal \Delta\boldsymbol{\omega}^2 \Delta\mathbfcal{R} =
-\frac{\mu^2}{2M} \boldsymbol{R}_0^\intercal [\boldsymbol{\omega}_\textrm{com}^2]^{-1} [\Delta\boldsymbol{\omega}^2]^2 \boldsymbol{R}_0.
\label{eq:mol-shift}
\end{equation}
This explains the shift of the molecular states seen in Fig.\ \ref{fig:direct-prod}. It shows that the shift is quadratic in the trap separation $z_0$ and is the same for all molecular states.

\begin{figure}
\includegraphics[width=0.43\textwidth]{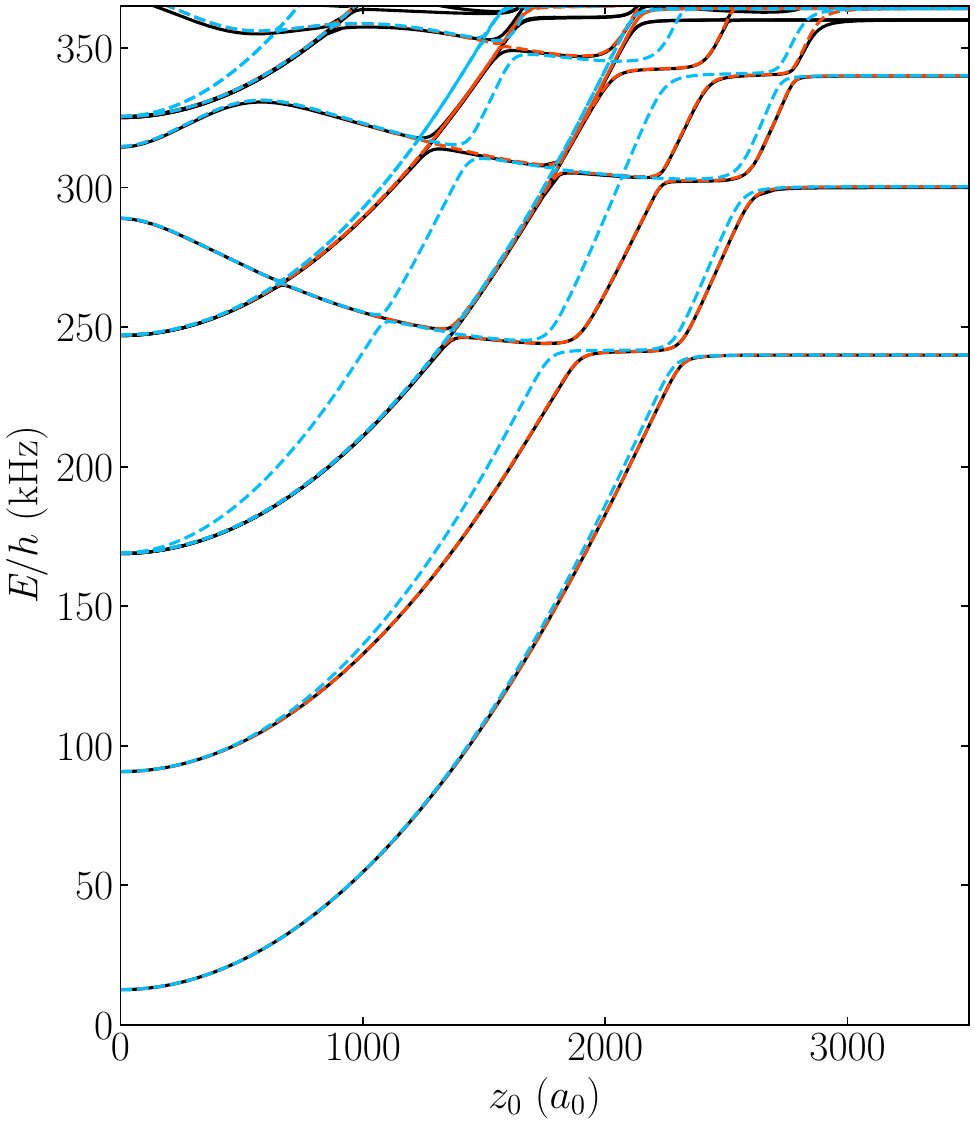}
\caption{Levels of Rb and Cs atoms in separated spherical traps, as in Fig.\ \ref{fig:direct-prod}, using different approaches. Black lines show results using a large shifted-molecule basis set (444)(444). Blue (or red) lines show results with smaller basis sets (444)(222) using the direct-product (or shifted-molecule) approach.}
\label{fig:direct-shifted-comp}
\end{figure}

Figure \ref{fig:direct-shifted-comp} compares results using small direct-product and shifted-molecule basis sets (with $(n_x^\textrm{max} n_y^\textrm{max} n_z^\textrm{max}) (N_X^\textrm{max} N_Y^\textrm{max} N_Z^\textrm{max}) = (444)(222)$) with those using a much larger shifted-molecule basis set (444)(444); the latter gives nearly converged results. The small shifted-molecule basis set gives very accurate results for all the molecular states and for the singly excited atom-pair states; its only visible deficiency in Fig.\ \ref{fig:direct-shifted-comp} is for the atom-pair states with $(n_\textrm{Rb},n_\textrm{Cs})=(0,2)$, which are unconverged with the smaller basis set of center-of-mass functions. The avoided crossings involving the ground and first-excited atom-pair states are all very accurately reproduced. The small direct-product basis set, by contrast, is substantially in error for several of the molecular states and their avoided crossings.

Comparison of Figs.\ \ref{fig:direct-prod}  and \ref{fig:direct-shifted-comp} demonstrates that even the (444)(444) basis set is significantly unconverged for the direct-product approach, producing unphysical non-degeneracies for both molecular and atom-pair states with larger values of $N_X$ and/or $N_Y$. The shifted-molecule approach performs much better in this respect; the (444)(222) basis set is adequate for most purposes, and contains only 972 functions, so that diagonalization is computationally cheaper by about a factor of 80. This basis set is used in the remainder of the paper, except where otherwise stated.

\begin{figure}[tbp]
	\includegraphics[width=0.43\textwidth]{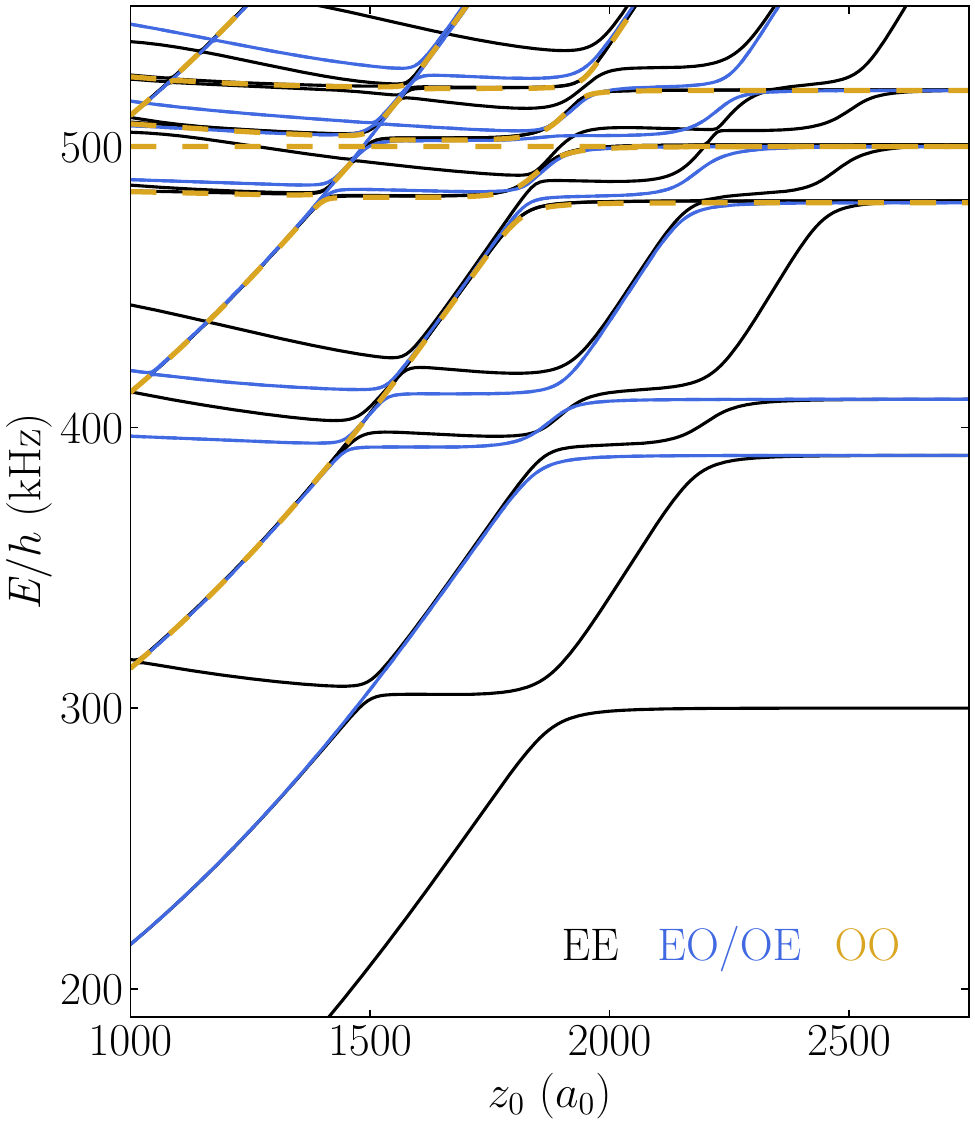}
	\caption{Levels of Rb and Cs atoms in separated spherical traps as a function of separation $z_0$, with $\omega_{\textrm{Rb}} = 110$ kHz and $\omega_{\textrm{Cs}} = 90$ kHz. Levels of EE, EO, OE, and OO symmetry are shown, but those of EO and OE symmetry are degenerate for spherical traps.}
\label{fig:levels-sym-sph}
\end{figure}

\section{Effects of coupling between relative and center-of-mass motion}
\label{sec:effects}

Figure \ref{fig:levels-sym-sph} shows the levels of different symmetries for Rb and Cs atoms in separated spherical traps with frequencies $\omega_{\textrm{Rb}} = 110$ kHz and $\omega_{\textrm{Cs}} = 90$ kHz. The levels of EE symmetry show complicated patterns of avoided crossings, which will be discussed further below. However, the levels of other symmetries are relatively simple. For spherical traps, the complete system has cylindrical symmetry, so levels of EO and OE symmetry are degenerate.

For spherical traps, the levels singly excited in either $\omega_{\textrm{Rb}}$ or $\omega_{\textrm{Cs}}$, with $n_\textrm{Rb}=1$ or $n_\textrm{Cs}=1$, are triply degenerate at large trap separations. Those with excitation along $x$ and $y$ have OE and EO symmetry, respectively. These two singly excited states show a $3\times 3$ avoided crossing with a molecular state near $1800\ a_0$ and a narrower one near $1500\ a_0$. When the traps are merged adiabatically an atom pair with single excitation in $\omega_\textrm{Cs}^{x(y)}$ (or more generally in the \emph{lower} of $\omega_1^{x(y)}$ and $\omega_2^{x(y)}$) will undergo mergoassociation to form a motionally excited molecule with $n_\textrm{com}^{x(y)}=1$. However, a pair with single excitation in $\omega_\textrm{Rb}^{x(y)}$ (i.e.\ in the \emph{higher} of $\omega_1^{x(y)}$ and $\omega_2^{x(y)}$) will be transferred to $n_\textrm{Cs}^{x(y)}=1$ at the two crossings near $1800\ a_0$; the pair may then pass either diabatically or adiabatically over the inner crossing; the former leaves the excitation in $n_\textrm{Cs}$, while the latter forms a molecule with $n_\textrm{com}^{x(y)}=1$ and $n_\textrm{com}^z=1$.

The lowest atom-pair state with OO symmetry has $n_\textrm{Cs}^x=1$ and $n_\textrm{Cs}^y=1$. An atom pair in this state can again undergo mergoassociation to form a motionally excited molecule, now with $n_\textrm{com}^x=1$ and $n_\textrm{com}^y=1$. However, replacing one or both excitations with $\omega_\textrm{Rb}^{x(y)}$ results in more complicated outcomes.

In the following, we focus on levels of EE symmetry, which are the most important for mergoassociation with well-cooled atoms.

\begin{figure*}[tbp]
	\includegraphics[width=\textwidth]{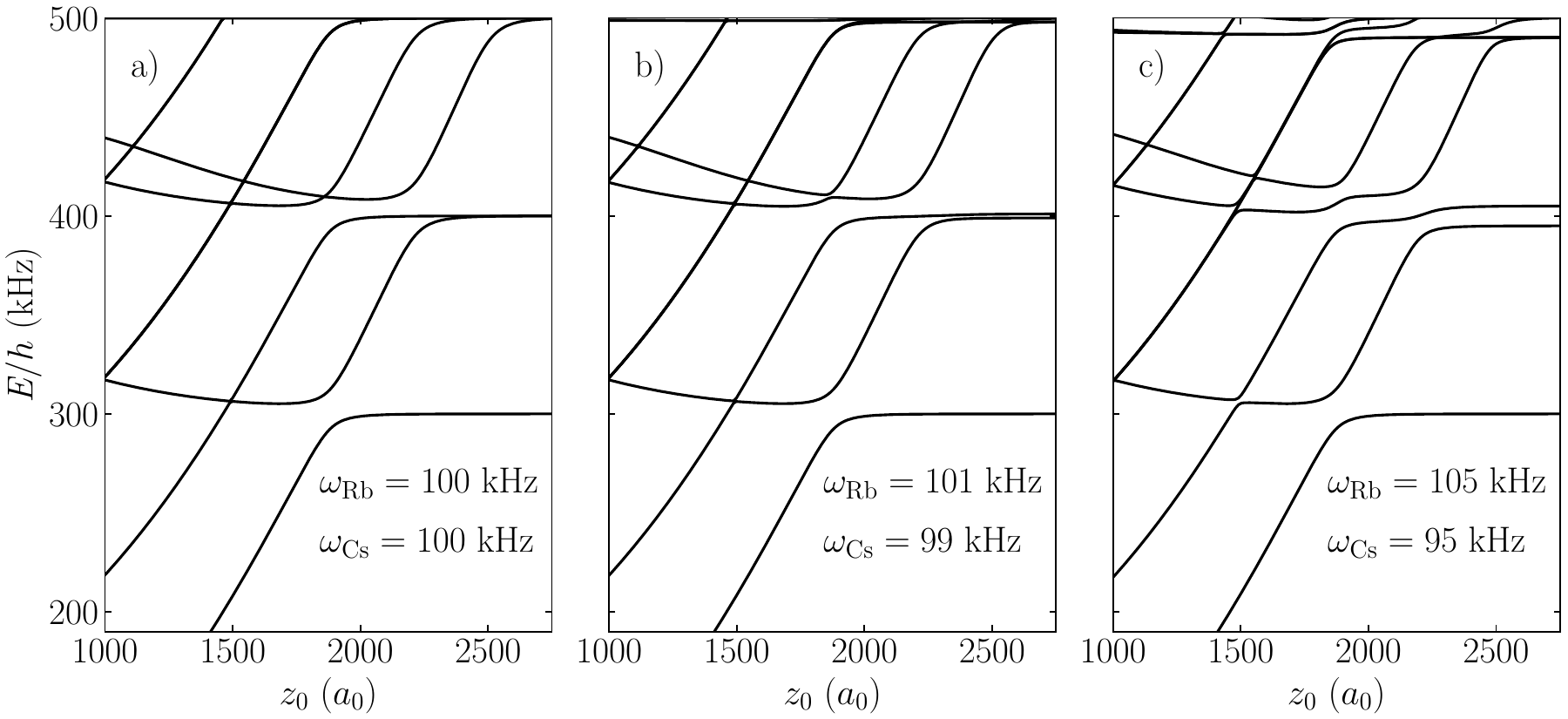}
	\caption{Levels of Rb and Cs in separated spherical traps as a function of separation $z_0$, with small differences between $\omega_1$ and $\omega_2$. Only levels of EE symmetry are shown.}
\label{fig:weak}
\end{figure*}

\subsection{Weak coupling}
\label{sec:weak}

When $\omega_1=\omega_2$, the relative and center-of-mass motions are completely uncoupled. The levels and avoided crossings for EE symmetry are then as in Fig.\ \ref{fig:weak}(a). The lowest atom-pair state shows an avoided crossing with the molecular state with no center-of-mass motion, but there is an unavoided crossing with the molecular state that is motionally excited. There are 2 atom-pair states with 1 unit of motional excitation. One of them may be viewed as excited in the relative coordinate but not the center-of-mass coordinate, so shows an avoided crossing with the molecular state with no center-of-mass motion but an unavoided crossing with the one that is motionally excited. The other may be viewed as excited in the center-of-mass coordinate, so shows an avoided crossing with the molecular state that is motionally excited but does not interact with the lowest molecular state.

Only a small difference between $\omega_1$ and $\omega_2$ is needed to change this picture. Figure \ref{fig:weak}(c) shows a crossing diagram with approximately 10\% difference between $\omega_1$ and $\omega_2$. Here the atom-pair states with 1 unit of motional excitation should be viewed at large separation as single-atom excitations for atom 1 and atom 2, respectively. This identification persists through the avoided crossings with both the ground and motionally excited molecular states. The molecular states, by contrast, remain best described as products of functions for relative and center-of-mass motion. Since the atom-pair states with excitation for a single atom are linear combinations of those with excitation in the relative and center-of-mass motions, there are strong avoided crossings between both atom-pair states and both molecular states.

Figure \ref{fig:weak}(b) shows an intermediate case with a 2\% difference between $\omega_1$ and $\omega_2$. Here the atom-pair states with 1 unit of motional excitation again correspond to single-atom excitations at very large $z_0$, but these states mix as the two traps approach one another. At the values of $z_0$ where the atom-pair states cross molecular states, $z_0^\textrm{X}$, this mixing is nearly complete and the levels are well described by quantum numbers for relative and center-of-mass motion. The diagram thus resembles Fig.\ \ref{fig:weak}(a): the lower singly-excited atom-pair state shows a strong avoided crossing with the molecular state with no center-of-mass motion, but a weak avoided crossing with the molecular state that is motionally excited. The situation is reversed for the upper singly-excited atom-pair state, which is mostly excited in the center-of-mass coordinate.

\subsection{Intermediate and strong coupling}
\label{sec:strong}

Figure \ref{fig:strong-crossings} shows level crossing diagrams for larger values of $\omega_\textrm{Rb}-\omega_\textrm{Cs}$. In this regime, the avoided crossings between atom-pair and molecular states are well isolated from one another and can each be characterized in terms of 2 interacting states.

\subsubsection{Mergoassociation with atoms in motional ground states}
\label{sec:crossing-A}

\begin{figure*}[tbp]
	\includegraphics[width=\textwidth]{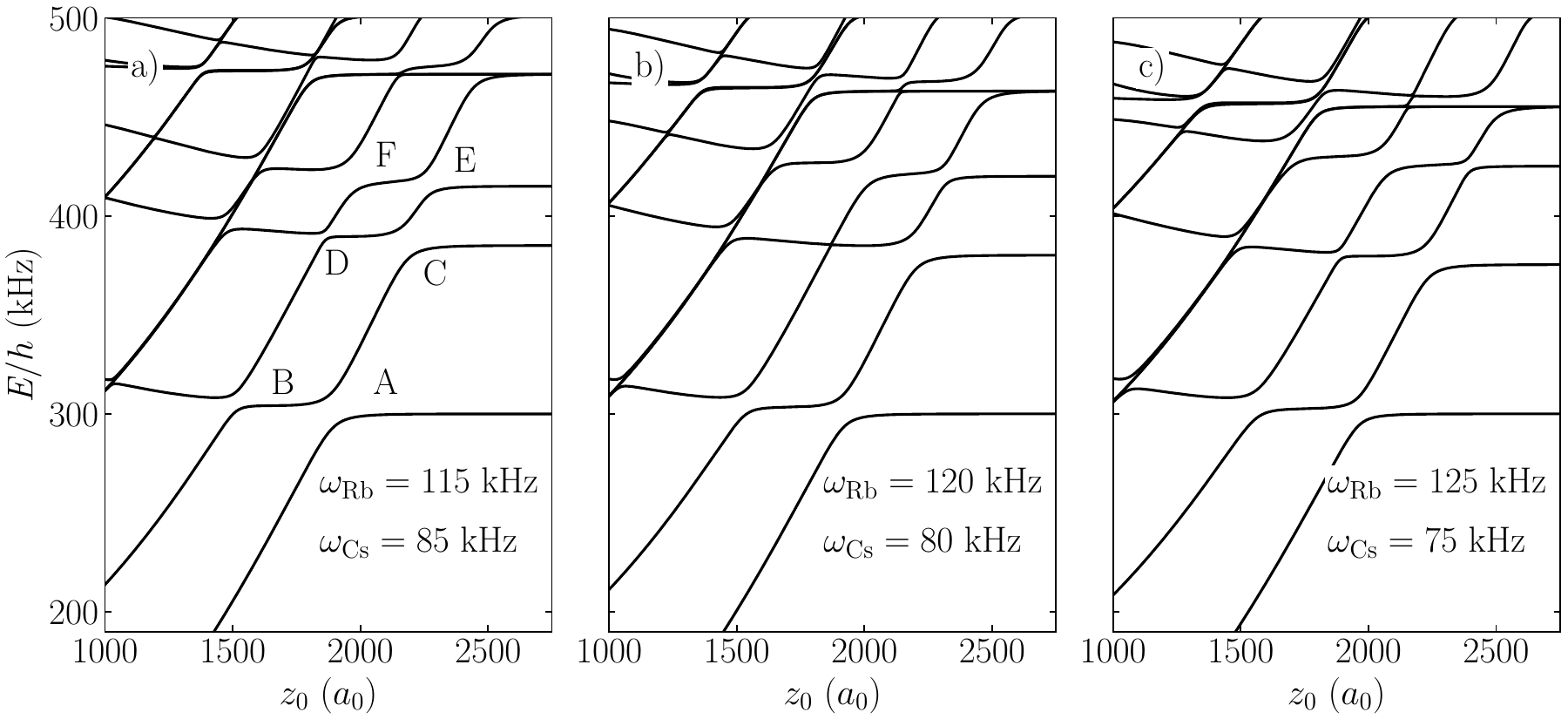}
	\caption{Levels of Rb and Cs in separated spherical traps as a function of separation $z_0$, with moderate differences between $\omega_1$ and $\omega_2$. Only levels of EE symmetry are shown.}
\label{fig:strong-crossings}
\end{figure*}

For mergoassociation from atoms in their motional ground states, the most important quantity is the strength $\Omega_\textrm{eff}$ of the lowest avoided crossing, near $z_0 = 2000\ a_0$ in Figs.\ \ref{fig:weak} and \ref{fig:strong-crossings}. The strength of this crossing for RbCs is shown in Fig.\ \ref{fig:crossing-strengths}(a) as a function of $\omega_\textrm{Rb}-\omega_\textrm{Cs}$. Here $\omega_\textrm{Rb}+\omega_\textrm{Cs}$ is held constant at 200~kHz, which keeps the energy of the lowest atom-pair state the same. However, the curvature of the molecular state is approximately proportional to $\omega^2_\textrm{rel}$, which is given by Eq.\ \ref{eq:wrel} and is not constant. As a result, the crossing distance $z_0^\textrm{X}$ generally increases as $|\omega_\textrm{Rb}-\omega_\textrm{Cs}|$ increases, though its minimum is slightly shifted from $\omega_\textrm{Rb}=\omega_\textrm{Cs}$. As shown in ref.\ \cite{Bird:mergo:2023}, the crossing strength depends principally on $\exp(-\frac{1}{2} z_0^\textrm{X}/\beta_\textrm{rel})$, so it decreases fast as $z_0^\textrm{X}$ increases; here $\beta_\textrm{rel} = (\hbar/\mu \omega_\textrm{rel})^\frac{1}{2}$. The dashed blue line on Fig.\ \ref{fig:crossing-strengths}(a) shows the result of the approximation from Eq.\ 55 of ref.\ \cite{Bird:mergo:2023}. The agreement is quite good, implying that the variation in $\Omega_\textrm{eff}$ with $\omega_\textrm{Rb}-\omega_\textrm{Cs}$ is dominated by the variation in $\omega_\textrm{rel}$ and hence in $z_0^\textrm{X}$, rather than by the coupling between relative and center of mass motions, characterized by $\Delta\omega^2$.

\subsubsection{Mergoassociation with motionally excited atoms}

\begin{figure}[tbp]
	\includegraphics[width=0.8\columnwidth]{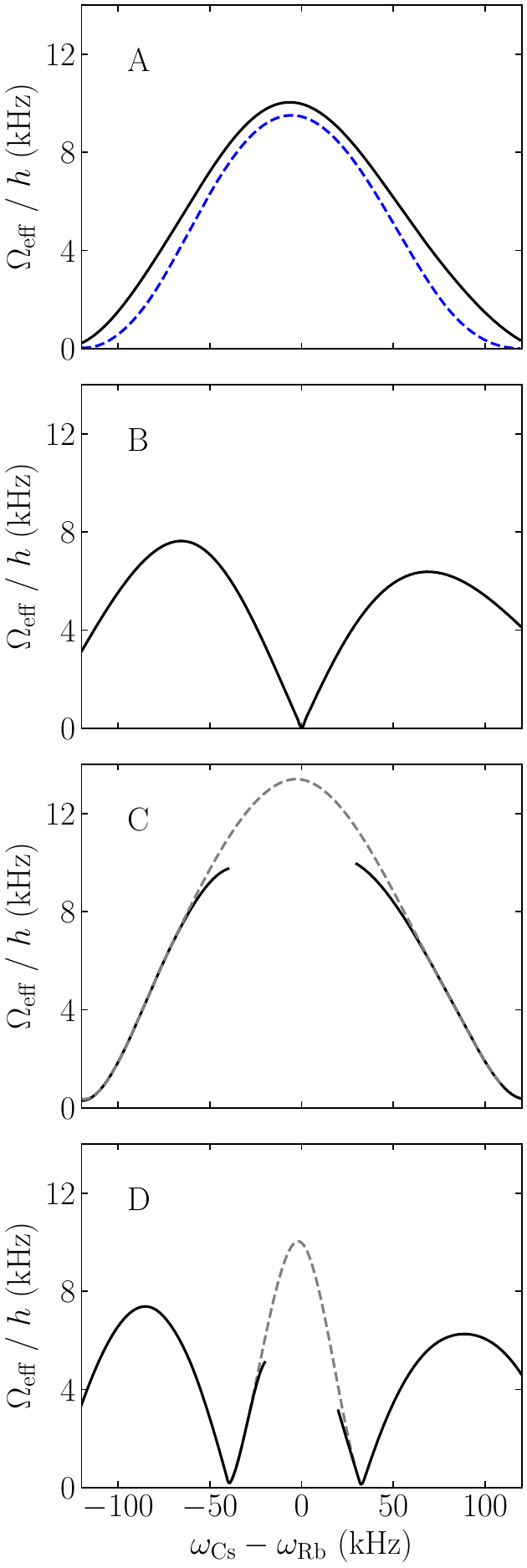}
\caption{The strength of avoided crossings A to D as a function of $\omega_\textrm{Cs}-\omega_\textrm{Rb}$, with $\omega_\textrm{Rb}+\omega_\textrm{Cs}$ held constant at 200 kHz. The black lines show the crossing strengths from the shifted-molecule approach. The grey dashed lines show interpolations through regions where a $2\times 2$ treatment breaks down, obtained as described in the text. The blue dashed line in (a) shows the result from Eq.\ 55 of ref.\ \cite{Bird:mergo:2023}.}
\label{fig:crossing-strengths}
\end{figure}

It is important to understand what happens when traps containing motionally excited atoms are merged. Under these circumstances, there are several avoided crossings that can be involved, labeled A to F in Fig.\ \ref{fig:strong-crossings}(a). The probability of traversing an avoided crossing adiabatically is quantified by the Landau-Zener formula, with a sufficiently slow merge producing adiabatic passage. The critical merging speed is proportional to $\Omega_\textrm{eff}^{-2}$ \cite{Bird:mergo:2023}.

There is interesting dependence of the strengths of the avoided crossings on $\omega_\textrm{Rb}-\omega_\textrm{Cs}$. As seen in section \ref{sec:crossing-A}, the strength of crossing A peaks near $\omega_\textrm{Cs}=\omega_\textrm{Rb}$. Conversely, the strength of crossing B, shown in Fig.\ \ref{fig:crossing-strengths}(b), is proportional to $|\omega_\textrm{Cs}-\omega_\textrm{Rb}|$ for small frequency differences; this arises because the relevant matrix element (Eq.\ \ref{eq:cpl-a-t}) includes a factor from the coupling between relative and center-of-mass motions. For larger frequency differences, the strength decreases for the same reasons as crossing A.

\begin{figure*}[tbp]
	\includegraphics[width=\textwidth]{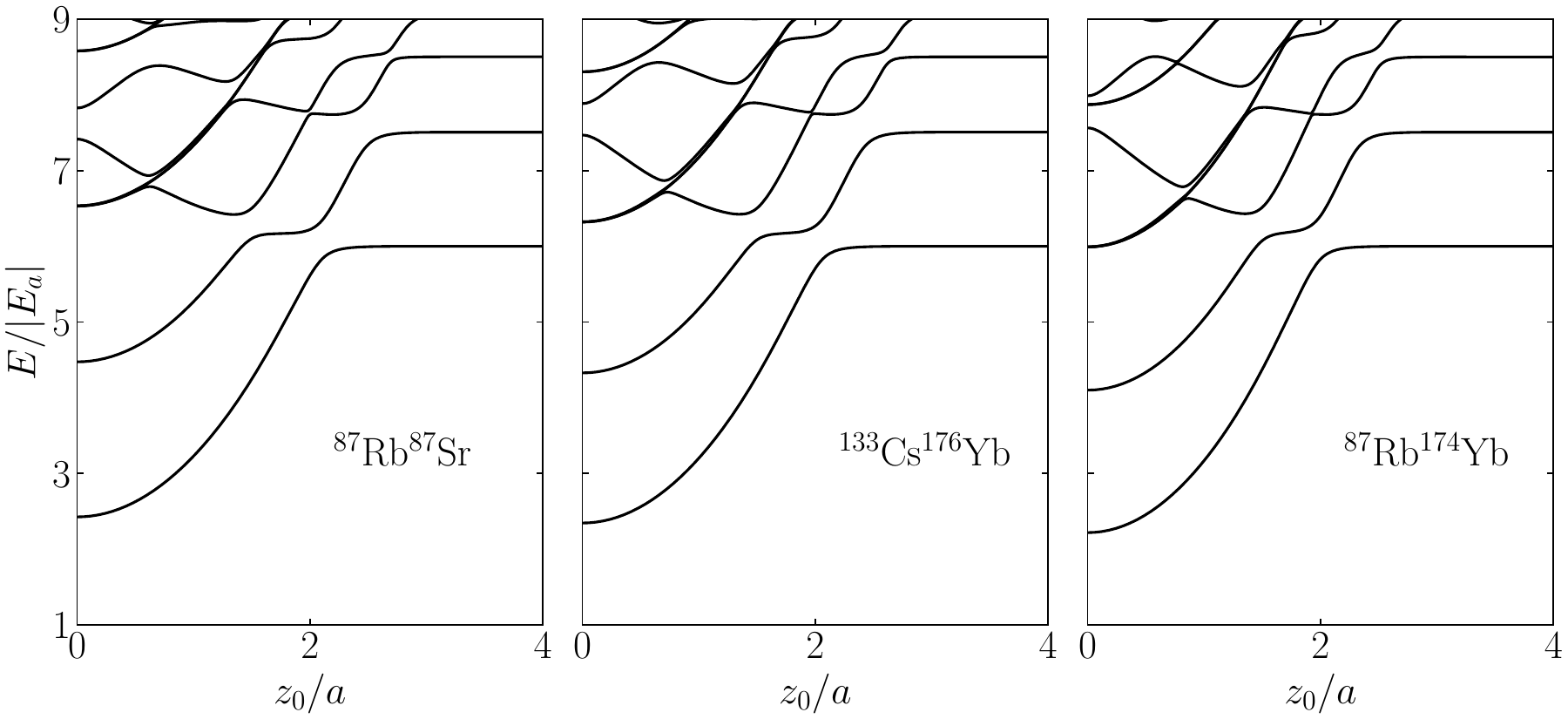}
	\caption{Level crossing diagrams for $^{87}$Rb$^{87}$Sr, $^{133}$Cs$^{176}$Yb and $^{87}$Rb$^{174}$Yb for $\hbar\omega_1 = 2.5 |E_a|$ and $\hbar\omega_2 = 1.5 |E_a|$. This corresponds to $(\omega_1,\omega_2) = (51.4,30.9)$, (94.0, 56.4) and (100.5, 60.3) kHz for the three systems, respectively.}
\label{fig:other-systems}
\end{figure*}

Crossings C, D, E, and F are more complicated. When $|\omega_\textrm{Cs}-\omega_\textrm{Rb}|$ is small, they involve the interaction of 3 states and do not lend themselves to simple characterization. This is again true when $\omega_\textrm{Rb}\approx 2\omega_\textrm{Cs}$ or $2\omega_\textrm{Rb}\approx \omega_\textrm{Cs}$, when the doubly excited state for one atom is close to the singly excited state for the other. Between these complicated regions, however, the crossing strengths may be characterized from a $2\times2$ model and are shown by the solid lines in Fig.\ \ref{fig:crossing-strengths}. The grey dashed lines show interpolations through the regions where a $2\times 2$ treatment breaks down; these are obtained by including a point at $\omega_\textrm{Rb}=\omega_\textrm{Cs}$, where the uncoupled problem can again be represented by a $2\times2$ matrix. The interpolations differ slightly from the solid curves in regions where a third state contributes significantly.

The separated atom-pair states involved in crossings C, D, E and F are characterized by quantum numbers $\{n_1,n_2\}=\{0,1\}$ and \{1,0\}, and may be approximately represented as linear combinations of $(n_\textrm{rel},n_\textrm{com})=(1,0)$ and (0,1). As a result, the matrix elements that govern their crossing strengths contain two terms, one proportional to $\omega_\textrm{Cs}-\omega_\textrm{Rb}$ and the other not. Because of this, the strengths of crossings D and E have minima due to destructive interference as a function of $\omega_\textrm{Cs}-\omega_\textrm{Rb}$; the minima are not actual zeroes, because \{0,1\} and \{1,0\} contain some contributions from states other than (1,0) and (0,1).

If the atom with the \emph{lower} trap frequency is motionally excited, it is possible to enter the state $(a,0)$ at avoided crossing C. From this point there are several possibilities. First, it may be possible to traverse crossing A diabatically with a fast merge, producing a molecule in state $(a,0)$ at small $z_0$. Alternatively, if crossing A is traversed adiabatically, the system will reach crossing B. For large frequency differences, crossing B can be traversed adiabatically, producing a motionally excited molecule in state $(a,1)$. For small differences, however, crossing B is very weak and is likely to be traversed diabatically, producing a ground-state atom pair. Yet another possibility is to pause the merging around $z_0\approx 2100\ a_0$ (for RbCs), which might allow optical transfer to a deeper state of the molecule.

If the atom with the \emph{higher} trap frequency is motionally excited, it is possible to enter the state $(a,0)$ at avoided crossing E. From this point there are many possible pathways based on different choices of adiabatic and diabatic traversals, controlled by merging speeds and trap frequencies. With a good understanding of the patterns of avoided crossings, it may be possible to devise sequences of merging and optical transfer that achieve efficient molecule formation even with motionally excited atoms.

\subsection{Mergoassociation for other systems}
\label{sec:other-systems}

Mergoassociation is potentially useful for many systems. As shown in ref.\ \cite{Bird:mergo:2023}, it is generally effective when the harmonic lengths of the traps or tweezers are comparable to (no more than a few times larger than) the scattering length. Otherwise, the lowest crossing occurs at large values of $z_0^\textrm{X}/\beta_\textrm{rel}$ and is too narrow to be useful. Mergoassociation is particularly promising for systems that lack Feshbach resonances, or where the Feshbach resonances are very narrow. Examples of this are systems of alkali-metal atoms with alkaline-earth atoms, where narrow resonances have been predicted \cite{Zuchowski:RbSr:2010, Brue:LiYb:2012, Brue:AlkYb:2013, Yang:CsYb:2019, Mukherjee:RbYb:2022} and observed \cite{Barbe:RbSr:2018, Franzen:CsYb-res:2022} but not yet used for magnetoassociation. RbSr, RbYb and CsYb all have isotopic combinations with large positive scattering lengths: $^{87}$Rb$^{87}$Sr with $a=1421(98)\ a_0$ \cite{Ciamei:RbSr:2018}, $^{87}$Rb$^{174}$Yb with $a=880(120)\ a_0$ \cite{Borkowski:2013} and $^{133}$Cs$^{176}$Yb with $a=798\ a_0$ \cite{Guttridge:2p:2018}.

In the absence of coupling between relative and center-of-mass motions, the mergoassociation problem scales conveniently with lengths expressed in terms of the relative harmonic length for relative motion, $\beta_\textrm{rel}$. This is the scaling we used in ref.\ \cite{Bird:mergo:2023}. However, in this representation, different scattering lengths $a$ produce a lowest avoided crossing for mergoassociation at different crossing distances $z_0^\textrm{X}/\beta_\textrm{rel}$. To compare systems with different $a$, it is more transparent to scale lengths according to $a$ and energies according to $|E_a|=\hbar^2/(2\mu a^2)$. In order to produce level crossing diagrams with molecular and atom-pair levels at approximately the same energies for different systems, we choose trapping frequencies that are the same multiple of $|E_a|$ for each system. This gives diagrams that are independent of $a$ and the mean atomic mass, but depend on the mass ratio $m_2/m_1$, which is close to 1 for RbSr, 1.3 for CsYb (compared to 1.53 for RbCs) and 2 for RbYb.

Figure \ref{fig:other-systems} shows level crossing diagrams for $^{87}$Rb$^{87}$Sr, $^{133}$Cs$^{176}$Yb and $^{87}$Rb$^{174}$Yb with $\hbar\omega_1 = 2.5 |E_a|$ and $\hbar\omega_2 = 1.5 |E_a|$. It may be seen that the crossing diagrams differ in detail, but show fairly similar patterns of avoided crossings in all the cases shown, with only weak dependence on the mass ratio. The one difference of any significance is that crossing D, near $z_0/a=2$ and $E/|E_a|=8$, is substantially stronger for RbSr than for the other systems, because the position of the minimum in Fig.\ \ref{fig:crossing-strengths}(d) depends on $m_2/m_1$. All three systems show substantial avoided crossings for pairs of atoms in a variety of motional states, so that mergoassociation is a promising method of molecule formation in all these systems.

\subsection{Trap anisotropy}
\label{sec:aniso}

\begin{figure}[tbp]
	\includegraphics[width=0.43\textwidth]{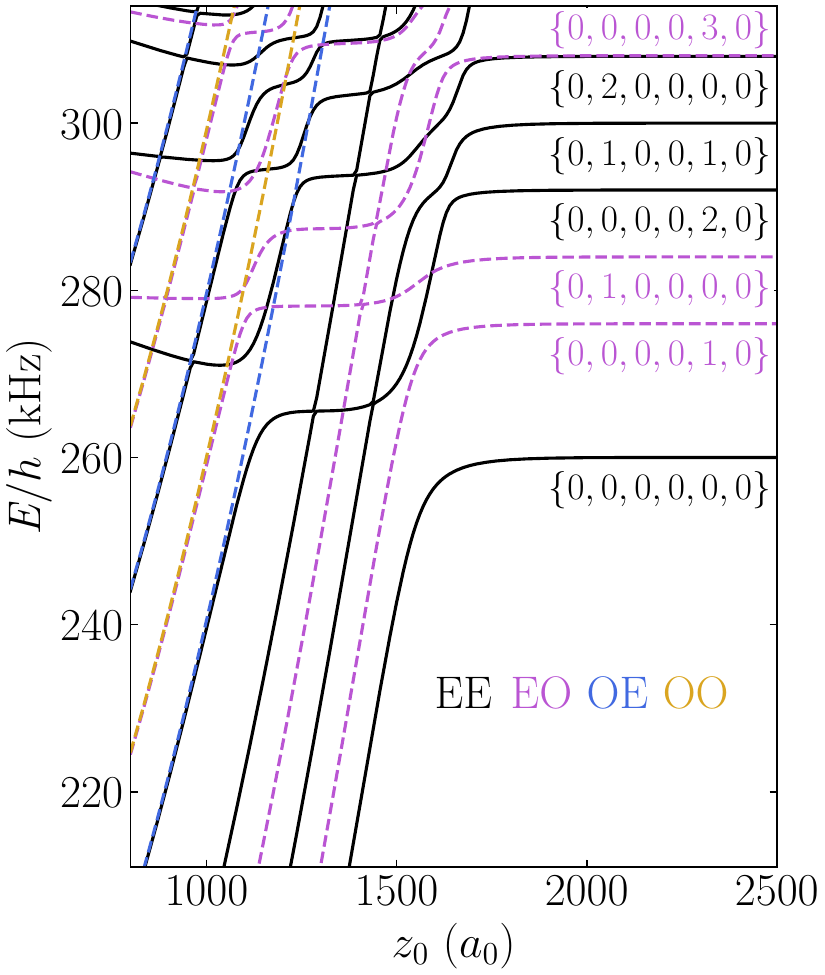}
	\caption{Levels of Rb and Cs atoms in separated anisotropic traps as a function of separation $z_0$, with \{$\omega_{\textrm{Rb}, x},\omega_{\textrm{Rb}, y},\omega_{\textrm{Rb}, z} \} = \{144, 24, 144\}$ kHz and \{$\omega_{\textrm{Cs}, x}, \omega_{\textrm{Cs}, y}, \omega_{\textrm{Cs}, z}\} = \{96, 16, 96 \}$ kHz. Levels with EE, EO, OE, and OO symmetry are shown. The atom-pair states are labeled with quantum numbers
$\{n_{\textrm{Rb},x},n_{\textrm{Rb},y},n_{\textrm{Rb},z},
n_{\textrm{Cs},x},n_{\textrm{Cs},y},n_{\textrm{Cs},z}\}$.}
\label{fig:levels-aniso}
\end{figure}

Optical tweezers are often strongly anisotropic, with much weaker confinement along the laser propagation axis than perpendicular to it. The mergoassociation experiments of Ruttley \emph{et al.}\ \cite{Ruttley:2023} were carried out on RbCs, using tweezers with frequency ratios $\omega_z/\omega_x \approx 1$ and $\omega_z/\omega_y \approx 6$ for both atoms. Figure \ref{fig:levels-aniso} shows the energy levels that result for a representative set of parameters, including both anisotropy and coupling between the relative and center-of-mass motions. All four symmetries are shown. The calculations used a shifted-molecule basis set with $(n_x^\textrm{max} n_y^\textrm{max} n_z^\textrm{max}) (N_X^\textrm{max} N_Y^\textrm{max} N_Z^\textrm{max}) = (444)(242)$. The details of the levels are complicated, and too specific to the individual case to justify detailed analysis here, but some important points are evident. First, the lowest crossing, involving atoms in their ground motional states, has a strength $\Omega_\textrm{eff} = 11.15$ kHz. This is similar to the strength obtained for spherical traps with frequencies chosen as $\omega_z$, which is $\Omega_\textrm{eff} = 11.84$ kHz. This justifies the spherical approximation used in ref.\ \cite{Ruttley:2023} to interpret the measured probabilities of diabatic and adiabatic crossing. If motional coupling is neglected, however, the crossing strength is 14.48 (15.61) kHz with anisotropy included (neglected). This demonstrates that the effects of motional coupling are significantly larger than those of anisotropy.

The avoided crossing involving the first-excited atom-pair state, with $n_{\textrm{Cs},y}=1$, is only slightly weaker than the lowest crossing. Atom pairs in this state may also be converted to molecules when the traps are merged. However, most other avoided crossings are substantially weaker. As noted above, the critical merging speed for adiabatic passage is proportional to $\Omega_\textrm{eff}^{-2}$; at the merging speeds used in ref.\ \cite{Ruttley:2023}, it is likely that these crossings would be traversed diabatically and fail to produce molecules.
%Analysing the outcome of a specific experiment, using Landau-Zener theory and its extensions, would be feasible in principle. However, such an analysis is not needed for the experiments of ref.\ \cite{Ruttley:2023} and is beyond the scope of the present paper.

\subsection{Logic gates}
\label{sec:logic}

Merging traps may also have applications in quantum information processing \cite{Stock:2003}. The interactions that control the energy levels depend on the hyperfine state of the atoms involved, so they may be used to accumulate phase differences between pairs of atoms in different states. This allows the production of controlled entanglement and the construction of 2-particle quantum-logic gates.

An important general insight from the present work is that, for a particular mass ratio $m_2/m_1$, the patterns of levels are ``universal" when lengths are expressed in terms of the scattering length and energies (and frequencies) are expressed in terms of the energy of the least-bound molecular state. Thus the key requirement for achieving differential phase shifts is that the scattering lengths are significantly different for the different pairs of atomic states involved. This is satisfied for most alkali-metal pairs, but not for all; modeling it requires a good understanding of the interaction potentials and detailed coupled-channel calculations using them \cite{Julienne:FD142, Takekoshi:RbCs:2012, Groebner:KCs:2017, Brookes:2022}.

The presence of coupling between relative and center-of-mass motions is a complicating factor for applications to logic gates. At the simplest level, such coupling modifies the trap separation at which the principal avoided crossing occurs, as described by Eq.\ \ref{eq:mol-shift}. It is important to take this into account. Nevertheless, for interactions involving pairs of atoms in their motional ground states, this is simply a quantitative correction.

Another issue is the feasibility (or fidelity) of quantum logic operations at finite temperature, when not all atoms are in their motional ground states. In this context, it would be desirable if the potential curves for motionally excited atoms were parallel to those for ground-state atoms. This occurs when the trapping frequencies for the two atoms are exactly equal, but not when the difference between them is significant. One possible advantage arises in cases where the difference is very small: then, as seen in Fig.\ \ref{fig:weak}(b), the potential curve for an atom excited in the \emph{higher} motional frequency is very similar to that for the absolute ground state, while that for an atom excited in the \emph{lower} frequency is not. Thus, if one atom is less well cooled that the other, it may be helpful to ensure that its trapping frequency is slightly (but as little as possible) \emph{higher} than that of its companion.

\section{Conclusions}
\label{sec:conclusions}

We have developed theoretical methods to calculate the energies of pairs of atoms in separated optical traps, taking account of both trap anisotropy and the coupling between relative and center-of-mass motions. The resulting levels are important both for molecule formation by mergoassociation and for potential applications in quantum information processing. We use basis sets based on Cartesian harmonic-oscillator functions for both relative and center-of-mass motion. The functions for relative motion are supplemented with a single molecular function. The effective trap potential for center-of-mass motion that is felt by the molecular function is shifted from the minimum of the combined trap; taking account of this shift complicates the algebra, but produces a substantial reduction in the size of the basis set needed for convergence.

Both mergoassociation and applications to quantum-logic gates rely on adiabatic passage over avoided crossings between atom-pair states and molecular states as a function of trap separation. The strengths of these avoided crossings are thus particularly important. We have used the example of RbCs to explore the dependence of the level patterns and the crossing strengths on the frequency difference between the traps for the two atoms. The lowest crossing, which is crucial for both applications, shifts to larger trap separations and becomes significantly weaker when center-of-mass motion is accounted for. Other crossings, which are important when merging traps containing motionally excited atoms, show more complicated behavior.

We have extended our treatment to other systems. Mergoassociation is generally feasible for atom pairs with positive scattering lengths that are comparable to or larger than the harmonic lengths of the traps. This corresponds to binding energies (for the least-bound state) that are not more than a few times the trap frequencies.  We have considered RbSr, RbYb and CsYb, which are resistant to magnetoassociation because their Feshbach resonances are so narrow and so sparse. All three systems have isotopic combinations with large positive scattering lengths. We have shown that, in units scaled by scattering lengths and binding energies, the level crossing diagrams are very similar for all three systems when the scaled trap freqencies are the same; they differ only because the ratio of atomic masses differs between systems. For all three systems, mergoassociation can form Feshbach molecules in the least-bound state with experimentally accessible trap frequencies.

Optical tweezer traps are usually strongly anisotropic, with much stronger confinement across the laser beam waist than along the beam. We have considered the combined effects of anisotropy and coupling between relative and center-of-mass motions for RbCs, using trap frequencies typical of current experiments. We have found that the effect of anisotropy is weaker than that of motional coupling under these conditions. We have also explored the effect of motional coupling on the levels that might be used for quantum-logic gates.
We have found that coupling between relative and center-of-mass motions can have substantial effects on the energy levels of separated traps. When merging traps containing atoms that are both in their motional ground states, the coupling leaves the general picture unchanged, but has significant effects that should be taken into account in quantitative work. However, when one or both atoms is in a motionally excited state, the coupling causes qualitative changes in the patterns of energy levels, which have important consequences for experimental outcomes.

\section*{Rights retention statement}

For the purpose of open access, the authors have applied a Creative Commons Attribution (CC BY) licence to any Author Accepted Manuscript version arising from this submission.

\section*{Data availability statement}

The data presented in this work are available from Durham University~\cite{DOI_data-merging-com}.

\section*{Acknowledgement}
We are grateful to Simon Cornish and Alex Guttridge for valuable discussions about the mergoassociation experiment.
This work was supported by the U.K. Engineering and Physical Sciences Research Council (EPSRC) Grant Nos.\ EP/P01058X/1, EP/T518001/1, EP/W00299X/1, and EP/V011677/1.

\begin{widetext}
\appendix
\section{Matrix elements}

The Hamiltonian used in the present work is
\begin{equation}
\hat{H} = \hat{T}_\textrm{rel}(\boldsymbol{R}) + V_\textrm{rel}^\textrm{trap}(\boldsymbol{R}) + V_\textrm{int}(\boldsymbol{R}) + \hat{H}_\textrm{com}^\textrm{trap}(\mathbfcal{R}) + V_\textrm{cpl}^\textrm{trap}(\boldsymbol{R},\mathbfcal{R}).
\end{equation}
The basis functions used here are products of functions in the relative coordinate $\boldsymbol{R}$ and functions in the center-of-mass coordinate $\mathbfcal{R}$.

\subsection{Relative motion}

For the relative coordinate, we use a nonorthogonal basis set formed from 3-dimensional harmonic-oscillator functions $| n_x n_y n_z \rangle = | n_x \rangle | n_y \rangle | n_z \rangle$, supplemented by a single molecular function $|a\rangle$. The harmonic-oscillator functions are
\begin{align}
\psi_{n}(\alpha) =
&= (2^n n! \beta_{\textrm{rel},\alpha})^{-1/2} \pi^{-1/4} H_n((\alpha-\alpha_0)/\beta_{\textrm{rel},\alpha}) \exp(-\textstyle{\frac{1}{2}} ((\alpha-\alpha_0)/\beta_{\textrm{rel},\alpha})^2),
\label{eq:harm-1d-fun}
\end{align}
where $\alpha=x$, $y$ or $z$, $\beta_{\textrm{rel},\alpha} = [\hbar/(\mu\omega_{\textrm{rel},\alpha})]^{1/2}$ and $H_n(q)$ is a Hermite polynomial. The corresponding eigenvalues are
\begin{align}
E_{n_x n_y n_z} = \hbar\omega_{\textrm{rel},x}(n_x+\textstyle{\frac{1}{2}})
+ \hbar\omega_{\textrm{rel},y}(n_y+\textstyle{\frac{1}{2}})
+ \hbar\omega_{\textrm{rel},z}(n_z+\textstyle{\frac{1}{2}}) .
\end{align}
For a contact potential, the molecular function $\psi_a = \langle R | a \rangle$ is given by Eq.\ \ref{eq:int-fun} and its eigenvalue by Eq.\ \ref{eq:int-e}.

The matrix elements for relative motion are as in ref.\ \cite{Bird:mergo:2023}.
The functions are normalized, so the diagonal elements of the overlap matrix ${\bf S}$ are all 1. The only non-zero off-diagonal elements of ${\bf S}$ are those between the molecular function and the harmonic-oscillator functions,
\begin{align}
S_{a,n_xn_yn_z} = \langle a | n_xn_yn_z \rangle =\int_0^{2\pi} \int_0^\pi \int_0^\infty \psi_a \psi_{n_xn_yn_z} R^2 dR\,\sin\theta d\theta \, d\phi.
\label{eq:S-int-harm}
\end{align}
These are evaluated by 3-dimensional numerical quadrature, using Gauss-Laguerre quadrature for $R$, Gauss-Legendre quadrature for $\theta$ and equally spaced and weighted points for $\phi$.

The elements of the Hamiltonian matrix for the harmonic-oscillator functions are
\begin{equation}
\langle n'_x n'_y n'_z | \hat{T}_\textrm{rel}(\boldsymbol{R}) + V_\textrm{rel}^\textrm{trap}(\boldsymbol{R}) | n_x n_y n_z \rangle = E_{n_xn_yn_z} \delta_{n'_x n_x} \delta_{n'_y n_y} \delta_{n'_z n_z},
\end{equation}
\begin{align}
\langle n'_x n'_y n'_z | V_\textrm{int}(\boldsymbol{R}) | n_x n_y n_z \rangle =
(2\pi\hbar^2 a /\mu)
\psi_{n'_x}(x_0) \psi_{n_x}(x_0) \psi_{n'_y}(y_0) \psi_{n_y}(y_0) \psi_{n'_z}(z_0) \psi_{n_z}(z_0).
\label{eq:Vint-harm}
\end{align}
For the molecular function,
\begin{align}
\langle a | \hat{T}_\textrm{rel}(\boldsymbol{R}) + V_\textrm{int}(\boldsymbol{R}) | a \rangle &= E_a;\\
\langle a | V_\textrm{rel}^\textrm{trap}(\boldsymbol{R}) | a \rangle &=
V_\textrm{rel}^\textrm{trap}(\boldsymbol{R}_0)
+\left( \frac{\mu a^2}{12}\right) (\omega_{\textrm{rel},x}^2 + \omega_{\textrm{rel},y}^2 + \omega_{\textrm{rel},z}^2).
\label{eq:a_Vtrap_a}
\end{align}
For a pure contact potential, $E_a=-\hbar^2/(2\mu a^2)$, but for real potentials this is accurate only for very large positive $a$ \cite{Julienne:Li67:2014}; when this approximation breaks down, it is best to choose $a$ to reproduce $E_a$, rather than vice versa.

The off-diagonal elements between the harmonic-oscillator functions and the molecular function are
\begin{align}
\langle a | \hat{T}_\textrm{rel}(\boldsymbol{R}) + V_\textrm{rel}^\textrm{trap}(\boldsymbol{R}) | n_x n_y n_z \rangle &= E_{n_xn_yn_z} S_{a,n_xn_yn_z}; \nonumber\\
\langle a | V_\textrm{int}(\boldsymbol{R}) | n_x n_y n_z \rangle &= -(\hbar^2/\mu) (2\pi/a)^{1/2}
\psi_{n_x}(x_0) \psi_{n_y}(y_0) \psi_{n_z}(z_0).
\label{eq:H-int-harm}
\end{align}

\subsection{Center-of-mass motion}

To include coupling to center-of-mass motion, we multiply each function in the basis set for relative motion with a set of 3-dimensional harmonic-oscillator functions in the center-of-mass coordinates, $|N_X N_Y N_Z \rangle = |N_X \rangle |N_Y \rangle |N_Z \rangle$, with functions $\Psi_\alpha(\alpha)$
defined by analogy with Eq.\ \ref{eq:harm-1d-fun}. The matrix elements of $\hat{T}_\textrm{rel}(\boldsymbol{R})$, $V_\textrm{rel}^\textrm{trap}(\boldsymbol{R})$, $V_\textrm{int}(\boldsymbol{R})$ and the overlap matrix are simply multiplied by overlaps between center-of-mass functions. In the direct-product approach, these are
\begin{align}
\langle N'_X N'_Y N'_Z | N_X N_Y N_Z \rangle =
\delta_{N'_X N_X} \delta_{N'_Y N_Y} \delta_{N'_Z N_Z}.
\end{align}
The matrix elements of $\hat{H}_\textrm{com}^\textrm{trap}$ are thus
\begin{align}
\langle n'_x n'_y n'_z N'_X N'_Y N'_Z | \hat{H}_\textrm{com}^\textrm{trap}(\mathbfcal{R}) | n_x n_y n_z N_X N_Y N_Z \rangle
&= E_{N_X N_Y N_Z} \delta_{n'_x n_x} \delta_{n'_y n_y} \delta_{n'_z n_z} \delta_{N'_X N_X} \delta_{N'_Y N_Y} \delta_{N'_Z N_Z}; \label{eq:com-t-t}\\
\langle a N'_X N'_Y N'_Z | \hat{H}_\textrm{com}^\textrm{trap}(\mathbfcal{R}) | a N_X N_Y N_Z \rangle
&= E_{N_X N_Y N_Z} \delta_{N'_X N_X} \delta_{N'_Y N_Y} \delta_{N'_Z N_Z}; \label{eq:com-a-a}\\
\langle a N'_X N'_Y N'_Z | \hat{H}_\textrm{com}^\textrm{trap}(\mathbfcal{R}) | n_x n_y n_z N_X N_Y N_Z \rangle
&= E_{N_X N_Y N_Z} S_{a,n_x n_y n_z} \delta_{N'_X N_X} \delta_{N'_Y N_Y} \delta_{N'_Z N_Z},
\label{eq:com-a-t}
\end{align}
where
\begin{align}
E_{N_X N_Y N_Z} = \hbar\omega_{\textrm{com},X}(N_X+\textstyle{\frac{1}{2}})
+ \hbar\omega_{\textrm{com},Y}(N_Y+\textstyle{\frac{1}{2}})
+ \hbar\omega_{\textrm{com},Z}(N_Z+\textstyle{\frac{1}{2}}) .
\end{align}
\subsection{Coupling between relative and center-of-mass motions}
The matrix elements of $V_\textrm{cpl}^\textrm{trap}(\boldsymbol{R},\mathbfcal{R})$ may be factorized
\begin{align}
\langle n'_x n'_y n'_z N'_X N'_Y N'_Z | V_\textrm{cpl}^\textrm{trap}(\boldsymbol{R},\mathbfcal{R}) | n_x n_y n_z N_X N_Y N_Z \rangle
&= \mu \langle n'_x n'_y n'_z | (\boldsymbol{R} - \boldsymbol{R}_0)^\intercal | n_x n_y n_z \rangle
\Delta\boldsymbol{\omega}^2
\langle N'_X N'_Y N'_Z | \mathbfcal{R} - \mathbfcal{R}_0 | N_X N_Y N_Z \rangle
\label{eq:cpl-t-t}\\
\langle a N'_X N'_Y N'_Z | V_\textrm{cpl}^\textrm{trap}(\boldsymbol{R},\mathbfcal{R}) | a N_X N_Y N_Z \rangle
&= \mu \langle a | (\boldsymbol{R} - \boldsymbol{R}_0)^\intercal | a \rangle
\Delta\boldsymbol{\omega}^2
\langle N'_X N'_Y N'_Z | \mathbfcal{R} - \mathbfcal{R}_0 | N_X N_Y N_Z \rangle; \label{eq:cpl-a-a}\\
\langle a N'_X N'_Y N'_Z | V_\textrm{cpl}^\textrm{trap}(\boldsymbol{R},\mathbfcal{R}) | n_x n_y n_z N_X N_Y N_Z \rangle
&= \mu \langle a | (\boldsymbol{R} - \boldsymbol{R}_0)^\intercal | n_x n_y n_z \rangle
\Delta\boldsymbol{\omega}^2
\langle N'_X N'_Y N'_Z | \mathbfcal{R} - \mathbfcal{R}_0 | N_X N_Y N_Z \rangle,
\label{eq:cpl-a-t}
\end{align}
where
\begin{align}
	\langle a | \boldsymbol{R} - \boldsymbol{R}_0 | a \rangle &= -\boldsymbol{R}_0.
\end{align}
Matrix elements involving $(\boldsymbol{R} - \boldsymbol{R}_0) | n_x n_y n_z \rangle$ are evaluated using the identity
\begin{align}
(z - z_0) | n_z \rangle &= 2^{-\frac{1}{2}} \beta_{\textrm{rel},z} \left( \sqrt{n_z}\, | n_z-1 \rangle + \sqrt{n_z+1}\, | n_z+1 \rangle \right),
\end{align}
and similarly for other components. Thus
\begin{align}
\langle a | z - z_0 | n_x n_y n_z \rangle &=
2^{-\frac{1}{2}} \beta_{\textrm{rel},z} \left(\sqrt{n_z}\, S_{a,n_xn_yn_z-1} + \sqrt{n_z+1}\, S_{a,n_xn_yn_z+1}\right) \\
\langle n'_x n'_y n'_z | z - z_0 | n_x n_y n_z \rangle &= \delta_{n'_x n_x} \delta_{n'_y n_y}
2^{-\frac{1}{2}} \beta_{\textrm{rel},z} \left( \delta_{n'_z,n_z-1} \sqrt{n_z} + \delta_{n'_z,n_z+1} \sqrt{n_z+1} \, \right),
\label{eq:zpzz}
\end{align}
with similar expressions for $x-x_0$, $y-y_0$. For the center-of-mass coordinates, the analogous expressions are
\begin{align}
\langle N'_X N'_Y N'_Z | Z - Z_0 | N_X N_Y N_Z \rangle &= \delta_{N'_X N_X} \delta_{N'_Y N_Y}
2^{-\frac{1}{2}} \beta_{\textrm{com},Z} \left( \delta_{N'_Z,N_Z-1} \sqrt{N_Z} + \delta_{N'_Z,N_Z+1} \sqrt{N_Z+1} \, \right),
\label{eq:ZpZZ}
\end{align}
and similarly for $X-X_0$ and $Y-Y_0$.

\subsection{Shifted-molecule basis set}

For the shifted-molecule basis set, the center-of-mass functions $|\tilde{N}_X \tilde{N}_Y \tilde{N}_Z \rangle$ are shifted in $\mathbfcal{R}$ for functions containing $|a\rangle$ but not for those containing $| n_x n_y n_z \rangle$. This leaves Eqs.\ \ref{eq:com-t-t} and \ref{eq:cpl-t-t} unchanged, but Eqs.\ \ref{eq:com-a-a} and \ref{eq:cpl-a-a} are replaced by
\begin{align}
\langle a \tilde{N}'_X \tilde{N}'_Y \tilde{N}'_Z | \hat{H}_\textrm{com}^\textrm{trap}(\mathbfcal{R}) + V_\textrm{cpl}^\textrm{trap}(\boldsymbol{R},\mathbfcal{R}) | a \tilde{N}_X \tilde{N}_Y \tilde{N}_Z \rangle
&= \left( E_{\tilde{N}_X \tilde{N}_Y \tilde{N}_Z} - \frac{\mu}{2} \boldsymbol{R}_0^\intercal \Delta\boldsymbol{\omega}^2 \Delta\mathbfcal{R} \right) \delta_{\tilde{N}'_X \tilde{N}_X} \delta_{\tilde{N}'_Y \tilde{N}_Y} \delta_{\tilde{N}'_Z \tilde{N}_Z}.
\end{align}
The matrix elements (\ref{eq:com-a-t}) and (\ref{eq:cpl-a-t}) are also modified because $|N_X N_Y N_Z\rangle$ and $|\tilde{N}_X \tilde{N}_Y \tilde{N}_Z\rangle$ are nonorthogonal,
\begin{align}
\langle a \tilde{N}'_X \tilde{N}'_Y \tilde{N}'_Z | \hat{H}_\textrm{com}^\textrm{trap}(\mathbfcal{R}) | n_x n_y n_z N_X N_Y N_Z \rangle
&= E_{N_X N_Y N_Z} S_{a,n_x n_y n_z} \langle \tilde{N}'_X | N_X \rangle \langle \tilde{N}'_Y | N_Y \rangle \langle \tilde{N}'_Z | N_Z \rangle;
\label{eq:com-mod-a-t} \\
\langle a \tilde{N}'_X \tilde{N}'_Y \tilde{N}'_Z | V_\textrm{cpl}^\textrm{trap}(\mathbfcal{R}) | n_x n_y n_z N_X N_Y N_Z \rangle
&= \mu \langle a | (\boldsymbol{R} - \boldsymbol{R}_0)^\intercal | n_x n_y n_z \rangle
\Delta\boldsymbol{\omega}^2
\langle \tilde{N}'_X \tilde{N}'_Y \tilde{N}'_Z | \mathbfcal{R} - \mathbfcal{R}_0 | N_X N_Y N_Z \rangle,
\label{eq:cpl-mod-a-t}
\end{align}
where the overlap integrals between shifted and unshifted functions along each Cartesian axis $\alpha$ are \cite{Gradshteyn:2015}
\begin{equation}
\langle \tilde{m} | n \rangle = \left(\frac{m!}{2^{n-m} n!}\right)^\frac{1}{2}  \rho_\alpha^{n-m} L_m^{n-m}(\rho_\alpha^2/2) \exp(-\rho_\alpha^2/4).
\label{eq:mod-overlap}
\end{equation}
Here $\rho_\alpha=\Delta\mathcal{R}_\alpha/\beta_{\textrm{com},\alpha}$, $n\ge m$ and $L_m^{n-m}$ is an associated Laguerre polynomial.

The matrix elements of $\mathbfcal{R} - \mathbfcal{R}_0$ in Eq.\ \ref{eq:cpl-mod-a-t} are expressed in terms of their Cartesian components,
\begin{align}
\langle \tilde{N}'_X \tilde{N}'_Y \tilde{N}'_Z | Z - Z_0 | N_X N_Y N_Z \rangle = \langle \tilde{N}'_X | N_X \rangle \langle \tilde{N}'_Y | N_Y \rangle 2^{-\frac{1}{2}} \beta_{\textrm{com},Z} \left( \langle \tilde{N}'_Z | N_Z-1 \rangle \sqrt{N_Z} + \langle \tilde{N}'_Z | N_Z+1 \rangle \sqrt{N_Z+1} \,\right)
\end{align}
and similarly for $X-X_0$ and $Y-Y_0$.

Finally, the non-zero off-diagonal elements of the overlap matrix are
\begin{align}
\langle a \tilde{N}'_X \tilde{N}'_Y \tilde{N}'_Z | n_x n_y n_z N_X N_Y N_Z \rangle &= S_{a, n_x n_y n_z} \langle \tilde{N}'_X | N_X \rangle \langle \tilde{N}'_Y | N_Y \rangle \langle \tilde{N}'_Z | N_Z \rangle.
\end{align}

\end{widetext}

\bibliography{../all,merging-tweezers-com-data}
\end{document}